\documentclass[a4paper,12pt,reqno,superscriptaddress,nofootinbib]{revtex4}

\usepackage[centertags]{amsmath}
\usepackage{amsfonts}
\usepackage{amssymb}
\usepackage{amsthm}
\usepackage{newlfont}
\usepackage{stmaryrd}
\usepackage{mathrsfs}
\usepackage{mathtools}
\usepackage{euscript}
\usepackage{graphicx}
\usepackage{enumerate}
\usepackage{color}
\usepackage[margin=1in]{geometry}

\usepackage{hyperref}
\usepackage{physics}
\usepackage[compatibility=false]{caption}
\usepackage[labelfont=bf,justification=RaggedRight,singlelinecheck=false]{caption}
\usepackage{subcaption}
\usepackage{siunitx}

\graphicspath{{./reset-plots/}}

\theoremstyle{plain}

\theoremstyle{definition}

\theoremstyle{remark}

\newcommand{\be}{\begin{equation}}
	\newcommand{\en}{\end{equation}}

\newcommand{\opunit}{\text{1}\kern-0.22em\text{l}}

\newcommand{\id}{\textrm{d}}

\DeclareMathAlphabet{\mathpzc}{OT1}{pzc}{m}{it}

\let\oldsqrt\sqrt
\def\sqrt{\mathpalette\DHLhksqrt}
\def\DHLhksqrt#1#2{%
	\setbox0=\hbox{$#1\oldsqrt{#2\,}$}\dimen0=\ht0
	\advance\dimen0-0.2\ht0
	\setbox2=\hbox{\vrule height\ht0 depth -\dimen0}%
	{\box0\lower0.4pt\box2}}

\let\be=\beta

\DeclareMathAlphabet{\mathpzc}{OT1}{pzc}{m}{it}

\def\bea{\begin{eqnarray}}
	\def\eea{\end{eqnarray}}
\def\ba{\begin{array}}
	\def\ea{\end{array}}

\begin{document}
	\title{Resetting photons}	
	\author{Guilherme Eduardo Freire Oliveira$^1$, Christian Maes$^2$ and Kasper Meerts$^{2}$\\ $^1$Departamento de Física, Universidade Federal de Minas Gerais \\ $^2$Instituut voor Theoretische Fysica, KU Leuven}

\begin{abstract}
Starting from a frequency diffusion process for a tagged photon which simulates relaxation to the Planck law, we introduce a resetting where photons lower their frequency at random times.
We consider two versions, one where the resetting to low frequency is independent of the existing frequency and a second case where the reduction in frequency scales with the original frequency.  The result is a nonlinear Markov process where the stationary distribution modifies the Planck law by abundance of low-frequency occupation. The physical relevance of such photon resetting processes can be found in explorations of nonequilibrium effects, e.g., via random expansions of a confined plasma or photon gas or via strongly inelastic scattering with matter.
\end{abstract}
\maketitle

%\tableofcontents
\section{Introduction}
Since its original conception in \cite{tong1,evans1,evans2,evans3}, resetting has been introduced and added to diffusion processes for a variety of reasons.  Typically, optimizing search strategies has been the underlying motivation, but one can also imagine physical resettings.  By the latter we mean the result of a time-dependent potential, for which there are random moments of confinement, or the random appearance in certain locations of attractors, or the random contraction/expansion of an enclosed volume.  In the present paper, we investigate a new scenario where resetting is applied in frequency space of photons.  In that way we explore physically motivated nonequilibrium effects on the Planck distribution, while also adding physical substance to the application of resetting for nonlinear diffusions, \cite{przem}.\\

Resetting photons to a lower frequency refers to reducing the wave vector (in the reciprocal lattice), which, in real space, refers to an expansion. One physical mechanism we can imagine here is that of a confined plasma where the confinement is continually lifted at random moments. On the other hand, repeated inelastic scatterings of photons inside a cavity with the electrons in the wall can also provide a source for resetting behavior. We can picture that photons instantaneously lose their energy to electrons, which is rapidly dissipated to an external bath.\\

Mathematically, the photon frequency resetting may take various forms: we focus on a resetting in terms of a Doppler shift where the frequency gets divided $\omega \rightarrow \omega/d$ by some number $d$ (divisor).  We can imagine however (and we will introduce) other resetting procedures corresponding to different physical mechanisms, but they yield pretty much the same effects, as we will see.  To avoid condensation of the photons at zero momentum, we apply a thermal push-back: when the tagged photon reaches zero frequency, it is redistributed following the Planck law. That may be due to other sources of black-body radiation but here is used only to conserve total photon number, creating a current in frequency space.  The  resetting is restricted to low-frequency values which is related to our major physics motivation; to explore possible deviations to the Planck law from random and abrupt expansion of a plasma or photon gas.   \\
To incorporate that resetting of photon frequency  $\omega$, we use a nonlinear Markov process for a tagged photon in a plasma where the main mechanism is Compton scattering. (Other radiation processes are easily added but are not considered here.)  The nonlinearity of the Markov process follows from the stimulated emission, a quantum feature, and the corresponding (nonlinear) Fokker-Planck equation is the well-known Kompaneets equation \cite{kompa}.  The latter describes relaxation to the Planck radiation law, $\propto \omega^3 / (\exp[\hbar\omega/(k_BT)]-1)$.    In certain cases however, e.g.\ for photons inside optomechanical cavities, while resetting  the frequency still makes sense, Compton scattering plays a minor role.  Still we stick for convenience to the setup where the radiation relaxes to thermal equilibrium within the context of the Kompaneets equation. For cosmological applications, e.g.\ in the primordial plasma, an erratic expansion of the universe has not been contemplated before (leading to a stochastic Doppler shift of photon frequency) but the Kompaneets equation is the standard tool there \cite{peebles}.\\

In the next section we recall the elements of the Kompaneets process (without resetting) in the context of elastic Compton scattering with thermal electrons. In Section \ref{res} we introduce the resetting mechanisms and their physical motivation.  The simulations are discussed in Section \ref{sim} and we obtain nonequilibrium photon distributions.  The abundance at low frequencies is not surprising, but interesting for understanding possible scenarios of breaking the Planck distribution of the cosmic microwave background, while preserving almost perfectly the moderate to larger frequency regime of the Planck law.  More (speculative) conclusions and possible applications are presented in the final Section~\ref{con}. 

\section{The Kompaneets process}

As reference process we consider a fluctuation dynamics which realizes the Kompaneets equation as its nonlinear Fokker-Planck equation. In fact, that process was recently introduced in \cite{paper2}, but here we briefly revisit the setup.

\subsection{Kompaneets equation}
Relaxation towards equilibrium of a photon gas in contact with a nondegenerate, nonrelativistic electron bath in thermal equilibrium at temperature $T$ can be achieved via Compton scattering. Starting from the semi-classical Boltzmann-Uehling-Uhlenbeck equation for a dilute plasma, Kompaneets arrived at an equation for the (average) occupation number $n(\tau,\omega)$ at frequency $\omega$ of the photon gas at time $\tau$, \cite{kompa}:
\begin{equation}\label{ke}
\omega^2\frac{\partial n}{\partial \tau}(\tau,\omega)= \frac{n_e\sigma_T 
	c}{m_e c^2}\frac{\partial }{\partial \omega}\omega^4\left\{k_B T 
\frac{\partial n}{\partial \omega}(\tau,\omega) + 
\hbar\left[1+n(\tau,\omega)\right]n(\tau,\omega)\right\}
\end{equation}
The constant $\sigma_T$ is the Thomson total cross section, and $n_e,m_e$ are  the density and mass of the electrons, respectively.
The $ mc^2 / k_BT$ multiplier to the timescale expresses that electrons are slow-moving and so the average collision barely shifts a photon's frequency.  That is what allowed Kompaneets to make a diffusion approximation to the Boltzmann-equation; see \cite{paper} for more details.  The $\omega^2$- dependence (hidden in the $\omega^4$)  is due to the fact that the frequency shift is proportional to the frequency itself, implying that the variance on the shift goes like $\omega^2$. 
The induced Compton scattering due to stimulated emission, \cite{liedahl, blandford}, is present in the nonlinearity of the drift term, which is the second term in \eqref{ke}.  Its origin is that the photon carries momentum  proportional to its frequency.  A more statistical-mechanical perspective and microscopic derivation of  \eqref{ke} is presented in \cite{paper}.\\

An important point is made by noting that the Kompaneets equation \eqref{ke} implements only the Compton interaction to write the dynamics for $n(\tau,\omega)$. That has its own approximations and restrictions. As it is usually the case with differential evolution equations in physics, if we include other mechanisms, such as Bremsstrahlung or double Compton interaction, their contribution is independent of the Compton interaction and appears only additively in \eqref{ke} (see e.g. \cite{kompa, lightman}). The same happens when adding resetting to low-enough frequencies, but we now think of Doppler shifts by metric or cavity expansion.\\

Stationarity is achieved when $n(\tau,\omega)$ becomes the Bose-Einstein distribution with chemical potential $\mu$,
\[
n_\text{BE}(\omega) = \frac{1}{e^{\beta(\hbar\omega-\mu)} - 1},\qquad \beta =\frac{1}{k_BT}
\]
The Kompaneets equation yields a good understanding of the dynamical origin of the cosmic microwave background and of the related Sunyaev-Zeldovich effect \cite{sunyaeveffect,sunyaev}.  Excellent reviews include \cite{practical,gui,zeldovich}. Extensions and generalizations are e.g. obtained in \cite{buet, pitrou,barbosa, brown, itoh, itoh2, cooper, kohyama1, kohyama2, kohyama3,paper}.\\

To have a dimensionless Kompaneets equation we use $x= \hbar \omega/k_B T$, to rewrite \eqref{ke} as
\begin{equation}\label{ake}
x^2\frac{\partial n}{\partial t}(t,x) = \frac{\partial }{\partial x}x^4\left\{
\frac{\partial n}{\partial x}(t,x) + 
\left[1+n(t,x)\right]n(t,x)\right\}
\end{equation}
The time-variable $\tau$ has also changed into the dimensionless Compton optical depth
\[
t = n_e\, \sigma_T\, c \frac{ k_B T }{m_e c^2} \, \tau\,\coloneqq \frac{ \tau}{\tau_C}
\]
There, $\tau_C$ is related to the Doppler shift, having
\begin{equation}\label{shift}
\left\langle\frac{1}{2\tau_c}\left(\frac{\Delta\omega}{\omega}\right)^2\right\rangle\approx \frac{ k_B T }{m_e c^2} n_e \sigma_T c \coloneqq \frac{1}{\tau_C}
\end{equation} 
for its variance, where the collision time $\tau_c = \ell/c$ is obtained from the mean free path of photons $\ell=(n_e\sigma_T)^{-1}$. In the expression above, the average is taken with respect to the distribution of electrons.\\

As it stands written the Kompaneets equation\eqref{ake} is expressed in function of the occupation number $n(t,x)$, which is a unitless quantity quite literally counting the average number of photons present in a given wave-mode. For our purposes, the equation must be written for the photon density $\rho(t,x)$ instead. This entails multiplying the occupation number by the density of states, followed by a normalization through division by the total number of photons $N$. The spectral probability density for photons in a box of volume $V$ with periodic boundary conditions will then equal (see \cite{paper2})
\begin{equation}\label{eq:spd}
\rho(t,x) = \frac{V}{N} \frac{1}{\pi^2} \left(\frac{k_B T}{\hbar c}\right)^3 
x^2 n(t,x)% = \frac{x^2 n(y,x)}{2\zeta(3) Z}
.\end{equation}
\\
In terms of the photon spectral density \eqref{eq:spd}, the Kompaneets equation \eqref{ake} becomes
\begin{equation}\label{kp}
\frac{\partial \rho}{\partial t} (t,x) = -\frac{\partial}{\partial x}\left[\left(4x- x^2\left(1+2\zeta(3) \,\frac{\rho(t,x)}{x^2}\right)\right)\rho(t,x)\right] + \frac{\partial^2}{\partial x^2}\left[x^2 \rho(t,x)\right]
\end{equation}
When $n(t,x)= n_\text{BE}(x) = 1 / (\exp(x) - 1)$ (Planck law) the photon number 
$N$ equals
\[
N_\text{BE} = 2 \zeta(3) \frac{V}{\pi^2} \left( \frac{k_B T}{\hbar c} \right)^3
\]
with $\zeta(3) = 1/2 \int_0^\infty \id x \,x^2/(e^x-1) \simeq 1.202$, and the spectral probability density is $ \rho_\text{BE}(x) = \frac{x^2\, n_\text{BE}(x)}{2\zeta(3)}$, the stationary equilibrium solution to \eqref{kp}.

\subsection{Tagged photon stochastic process}
The Kompaneets equation is positivity preserving, \cite{positivity}. It allows therefore a probabilistic interpretation as nonlinear Fokker-Planck equation. That was explicitly realized in \cite{paper2}, and we refer to the corresponding stochastic dynamics of a tagged photon as the Kompaneets process. \\
A more general  tagged photon diffusion process is, in It\^o-convention,
\begin{equation} \label{kp-ito2}
\dot x	= B(t,x,n_t)   + \sqrt{2{D}(x)}\, \xi_t
\end{equation}
where $\xi_t$ is standard white noise and
\[
B(t,x,n_t)=2\frac{{D}(x)}{x} + D'(x) - \beta{D}(x) {U'}(x)(1+n(t,x))
\]
is the drift term.  Note, from the last term, that we need to know the particle number $n(t,x)$ to update the dimensionless frequency $x$ following \eqref{kp-ito2}.  %to use the relation \eqref{eq:spd} and thus find the occupation number $n(t,x)$ from the density $\rho(t,x)$.   In our case, particle number is conserved.\\
The It\^o--stochastic process \eqref{kp-ito2} is our Kompaneets process in the case where
\[
\beta U(x) = x,\qquad D(x) = x^2
\]
Note also that $x\geq 0$. 
Even without explicit boundary conditions, there is however no probability flux through the origin $\omega=0$, meaning that negative frequencies are not observed in the simulations. We therefore in the present work ignore the detailed implementation of radiative processes such as {\it Bremsstrahlung} or double Compton scattering that in real nature control the photon number density, picking up and absorbing photons with low-enough frequencies.\\
  
\begin{figure}
	\includegraphics[width=0.8\textwidth]{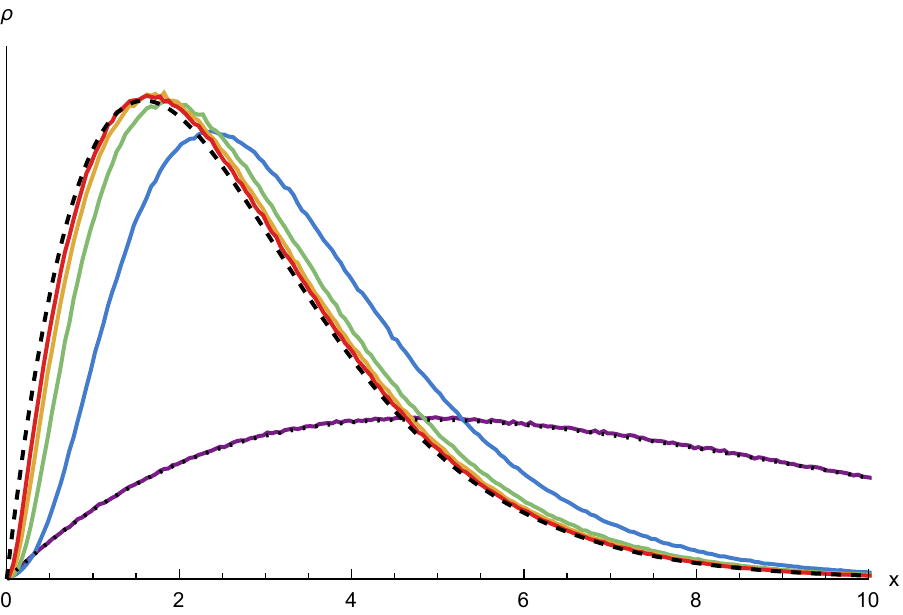}
	\caption{Frequency distribution for the Kompaneets process for $t=0.0,0.5,1.0,1.5$ and $2.0$. The initial condition is a ``hot Planck Law" at a temperature three times the temperature $T$ used in the dynamics, corresponding to the rescale $x\to x/3$ of the dimensionless frequency. There is a fast pre-thermalization for the moderate and high frequencies, followed by a slower relaxation at the lower frequency scale.  See more details in \cite{paper2}.}
	\label{fig:spd-kompaneets}
\end{figure}

As an illustration of the soundness of the process, we reproduce here in Fig.\ref{fig:spd-kompaneets} the results obtained from the simulation of the stochastic equation \eqref{kp} using the Euler-Maruyama algorithm \cite{toral}. To implement stimulated emission, which makes a nontrivial aspect both in the simulation and theory, we consider an ensemble of $N$ processes, using the empirical histogram of frequencies to update the drift term at each timestep accordingly. The details of the implementation including reactive mechanisms together with a more comprehensive discussion can be found in \cite{paper2}. From Fig.~\ref{fig:spd-kompaneets} we see the relaxation towards the Planck law, confirming the validity of the implementation scheme.
  
\section{Resetting the Kompaneets process}\label{res}

We start from the Kompaneets process \eqref{kp-ito2} defined in the previous section and add Poissonian resetting with 
a constant rate. The resetting rate is denoted by $r>0$. Three methods are 
introduced for resetting, the so-called ``division'' method, where the energy of 
the photon is divided by a constant factor $d$, the ``uniform'' method, where the 
photon's energy is reset to a random value uniformly distributed between 0 and 
a cutoff $x_0$, and finally the ``exponential'' method, where the new energy of 
the photon follows an exponential distribution with scale $1/\lambda_0$.\\
Similar choices of resetting positions have been considered before; see for example \cite{evans2, solis-salas}.\\

For a justification of introducing such a resetting of the photon frequency, we suggest two possibilities.  A first one is imagining a discontinuous expansion of the Universe, on top of the usual and smooth expansion predicted by the highly symmetric Friedmann–Lemaître–Robertson–Walker (FLRW)--solution of the Einstein field equations.  From the beginning of the lepton epoch (where leptons dominated the mass of the Universe) to the time of recombination (when photons decoupled from matter), space has 
expanded a millionfold, and from recombination until now the scale factor has 
increased by another factor of one thousand. In the FLRW--solution, this scale factor has risen 
continuously and homogeneously. Moving away from this highly idealized FLRW--scenario, we can imagine indeed more localized, abrupt increases, applying only to a 
fraction of photons.  As a photon's frequency is inversely proportional to the 
scale factor, that in effect is a division of photon 
frequencies by some large factor at random times.  The process is reminiscent of the Doppler shift in 
frequency, and in fact mathematically and conceptually that is identical to that of Doppler shifts from expanding matter.  The metric expansion of space and the Doppler 
effect amount to the same thing concerning frequency shift. \\
 In this 
regard a second motivation exists, which might be experimentally feasible.  There we suggest to consider a cavity with randomly, quickly receding walls.  Any photon catching up to these walls would again undergo a downwards shift in  frequency.  It would be interesting to see how the nonequilibrium frequency distribution  starts to deviate from the Planck Law.  Interestingly, Fermi in 1949, \cite{fermi}, considered the opposite scenario for understanding the origin of high-energy cosmic rays.  He suggested that relativistic particles would accelerate by means
of their collision with interstellar clouds that move randomly and act as ’magnetic mirrors’.  The average energy gain is proportional there to $(v_c/v_p)^2$ where $v_c$ is the speed of the cloud and $v_p$ the speed of the particle.  That is similar to the loss in energy of our photons due to the low-frequency resetting.\\
Related to all that is our reason to reset to lower frequencies only; we look at abrupt expansion of the plasma or gas.  Yet, as a matter of fact, if we were to reset the photons to larger frequencies, the average frequency could be much larger than $k_BT$, which disagrees with the assumptions needed for the derivation of the Kompaneets equation. That is not a problem in our case, since we reset photons to lower frequencies.

The above suggests that the frequency of the photon be updated by the following 
stochastic rules: in an infinitesimal interval $\dd t$ we have
\begin{align}\label{eq:resetted-ito}
x(t + \dd t) &= x(t) + B(t,x,n_t) \dd t + \sqrt{2{D}(x)\dd t}\, \xi_t &&\text{with probability $1-r\dd t$}\\\nonumber
&= \begin{cases}x(t) / d &\text{``division'' method}\\
x\sim \operatorname{Uniform}(0,x_0)&\text{``uniform'' method}\\
x\sim \operatorname{Exp}(\lambda_0)&\text{``exponential'' method}\\ 
\end{cases}&&\text{with probability $r \dd t$}
\end{align}

The resetting protocol, therefore, requires the specification of two parameters: the resetting rate $r$, which controls the strength of resetting, i.e., larger $r$ produces larger population of reset photons in the stationary distribution; and the protocol parameter $\{d,x_0,\lambda_0\}$, which controls the range in frequency space where the reset photon is placed. In that sense, the protocol parameter, acts similarly to a ``cutoff''.  It is natural to choose $r \leq 0.1$ as the Kompaneets process has itself a relaxation time of order one, which is one order faster then. 
It is also observed in the present work that the effects of resetting are negligible in the stationary distribution when $r\lessapprox 10^{-3}$ for the range of other parameters considered here.\\

We avoid condensation of photons at zero frequency by a thermal push-back.  In other words, whenever the tagged photon reaches zero frequency, it is getting a new frequency distributed according to the Planck law.  In that way, we create a frequency current, where the resetting drives to lower frequency and the thermal push-back induces higher frequencies (when needed).  As physical realization of that thermal push-back we may imagine other sources of black-body realization that supplement the photon gas to keep the same photon number and reach a stationary (nontrivial) distribution.\\

As a final remark, we emphasize that most studies of stochastic resetting consider linear diffusion processes as they allow analytic treatments. Our diffusion process is nonlinear  as even the Fokker-Planck equation is nonlinear in the probability density, and on a fundamental level.  So far, \cite{przem} is the only other reference in the literature. In our case the nonlinearity has a specific physical origin which is the (quantum feature of) stimulated emission (for photons).   In the above, we have chosen for a fluctuation approach, allowing simulation.  In principle, we could also consider a perturbative approach.  In fact, the linearized Kompaneets equation considered as a regular Fokker-Planck equation has a kernel which is exactly known; see \cite{nagirner}. While that linear-Kompaneets scenario can be regarded as a valid regime when considering very low temperatures for the electron bath, we deal with a case where the nonlinearity is so strong that the perturbative approach would not be so relevant.  At any rate, simulation is also giving a more robust method.  After all,  resetting is an idealization  and as usual in physics and their experiments, validity and relevance are functions of energy, length- and time-scales, which, here and elsewhere, is important to realize in moving from theory to experiment.

\section{Simulation results and discussion}\label{sim}

To simulate these dynamics, we revisit the procedure from \cite{paper2}. We consider an ensemble of \num{e7} particles, which are binned in frequency space with a width of \num{e-3}. We run the simulation until $t = 10$, with a timestep $\Delta t$ of \num{e-3}. Adding to this, at every timestep and for every particle, we perform the resetting step with a probability $r \Delta t$; see \eqref{eq:resetted-ito}. The timestep $\Delta t$ is chosen such  that this product is much smaller than 1, making sure that the probability of two resets happening to the same particle in the same timestep remains negligible.\\

\begin{figure}
	\includegraphics{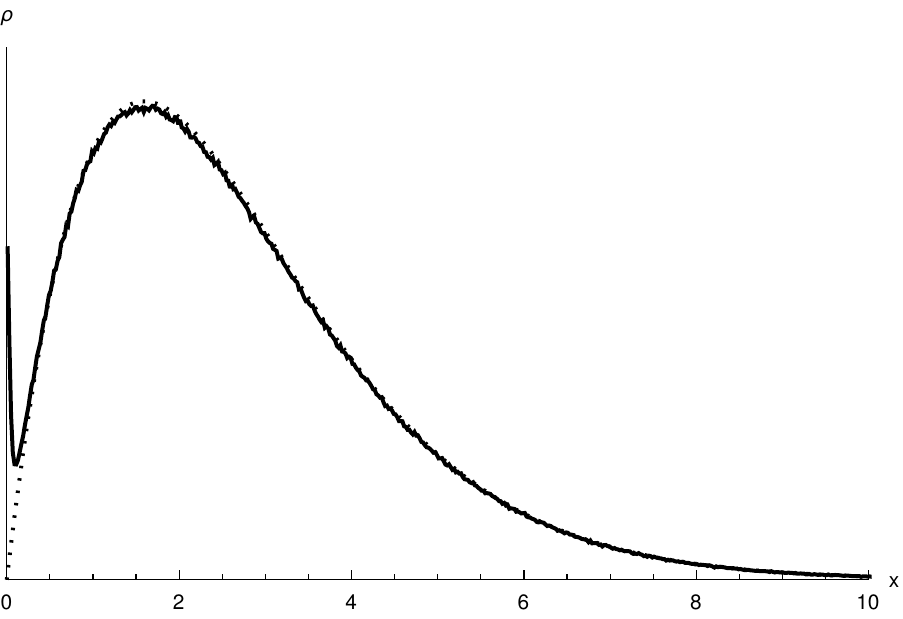}
	\caption{Stationary distribution for a resetting rate of $r=$\num{e-2} and a divisor $d=$\num{100}. For comparison, the dotted line gives the Planck distribution}\label{fig:fullplanck}
\end{figure}

\begin{figure}
	\subcaptionbox{$r=0.02$}{\includegraphics[width=0.48\textwidth]{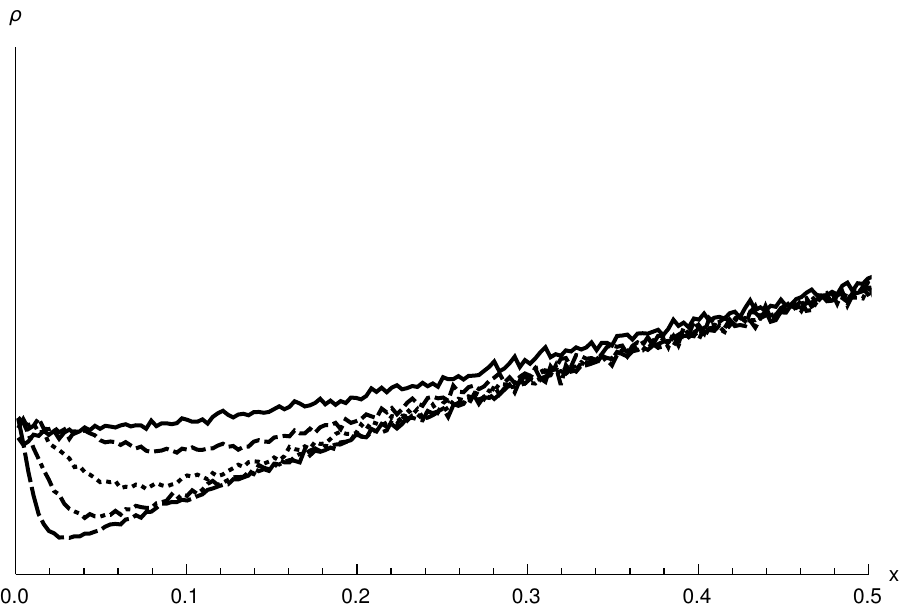}}\hfill%
	\subcaptionbox{$r=0.05$}{\includegraphics[width=0.48\textwidth]{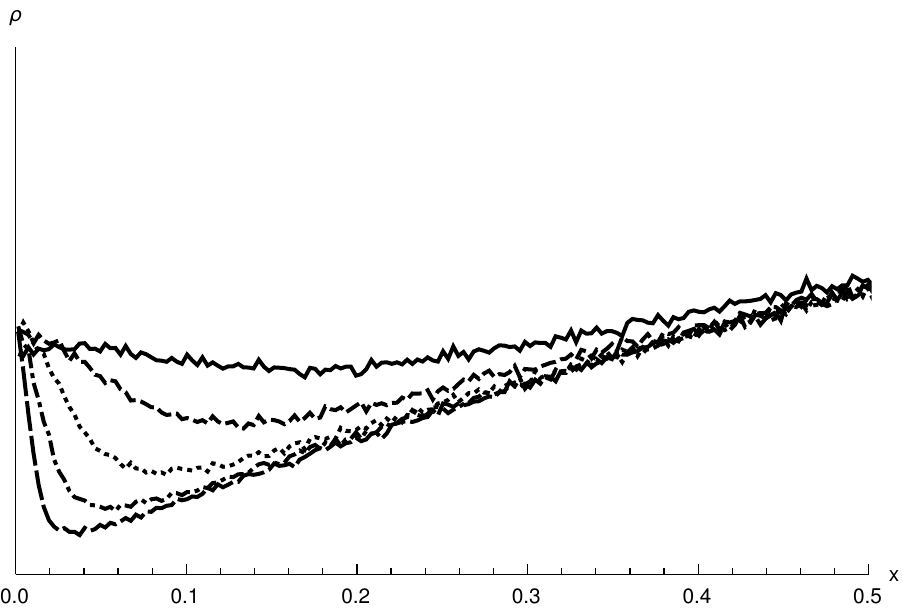}}
	\subcaptionbox{$r=0.10$}{\includegraphics[width=0.48\textwidth]{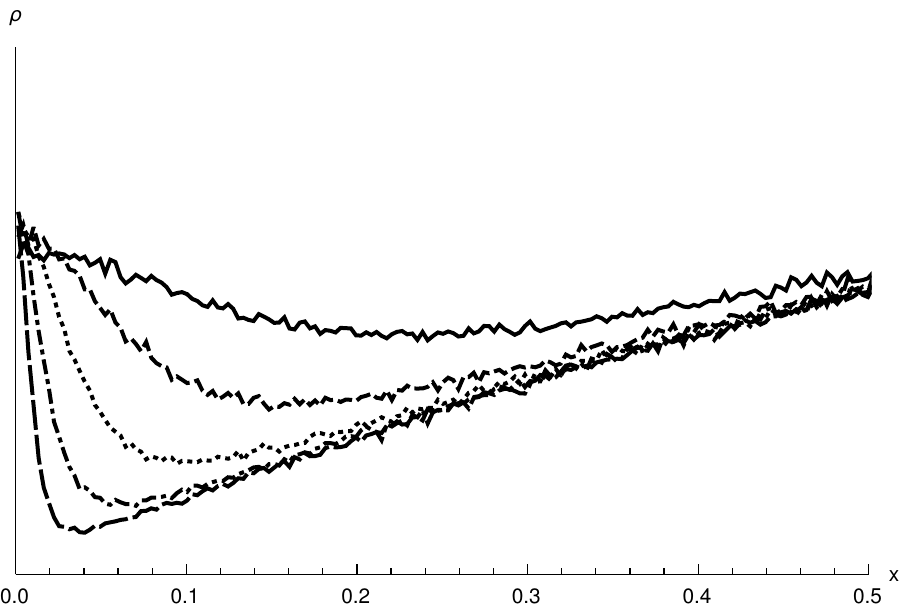}}\hfill%
	\subcaptionbox{$r=0.20$}{\includegraphics[width=0.48\textwidth]{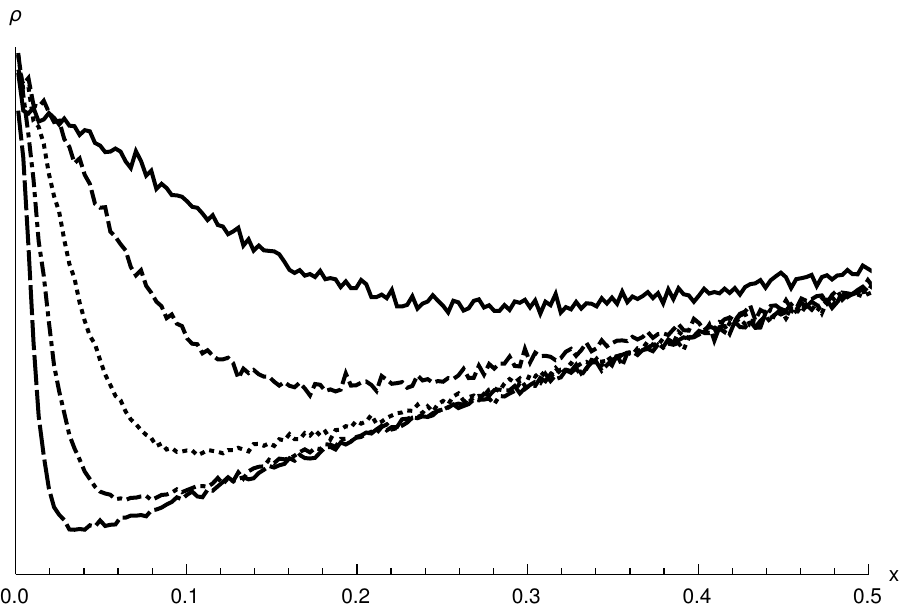}}
	\caption{Stationary distributions at low frequencies for four resetting rates $r$.  In each case, as the curves move down, the divisor $d =25, 50, 100, 200, 400$.}\label{fig:exploration}
\end{figure}

The results of such a simulation are shown in Fig.~\ref{fig:fullplanck}. With the average time between resets for a given photon being much smaller than the characteristic timescale of the Kompaneets process, it comes to no surprise that the deviations from the Planck law are mostly contained around the origin. After all, as we saw also in Fig.~\ref{fig:spd-kompaneets}, relaxation to the Planck law takes a time of order 1, while the resetting rate $r\leq 0.1$. In Fig.~\ref{fig:fullplanck} we see little or no difference between the resetting and no-resetting for reaching the Planck distribution at moderate to large frequencies. Therefore, in Fig.~\ref{fig:exploration}, we zoom in to the lower frequencies (where relaxation in the Kompaneets process is slowest), showing $x$ only from $0$ to $0.5$ (which is about 5\% of the full frequency range). From that figure, we clearly observe deviations from the Planck law in that frequency range due to resetting.\\

A first aspect to observe there, is that the different resetting procedures in \eqref{eq:resetted-ito} yield similar qualitative behavior by tuning properly the parameters $\{d,x_0,\lambda_0\}$.  That happens already by simply tuning the average photon frequency under the resetting distribution to match for the different resetting protocols. According to our definitions for the ``uniform'' and ``exponential'' methods we have the following averages after many resetting events,
\begin{align*}\langle x \rangle_{\text{Uni}} = \frac{x_0}{2} &&\text{and}&& \langle x \rangle_\text{Exp} = \frac{1}{\lambda_0}\end{align*}
while for the ``division'' protocol, because the timescale of resetting is smaller than the timescale of the relaxation towards the Planck law, we can to a high degree of accuracy assume the photon right before resetting to be distributed according to the Planck distribution, which gives it an average frequency of
\[\langle x \rangle_\text{Planck} = \int_0^{\infty}\dd x\,  x\rho_\text{Planck}(x) \approx 2.701...\]
where the Planck distribution is $\rho_\text{Planck}(x) = \frac{1}{2\zeta(3)}\frac{1}{e^{x}-1}$. Then after many ``division'' resetting events, the expected photon frequency is
\[\langle x\rangle_{\text{Div}}=\frac{\langle x \rangle_\text{Planck}}{d} \]

\begin{figure}[th]
	\includegraphics[width=\textwidth]{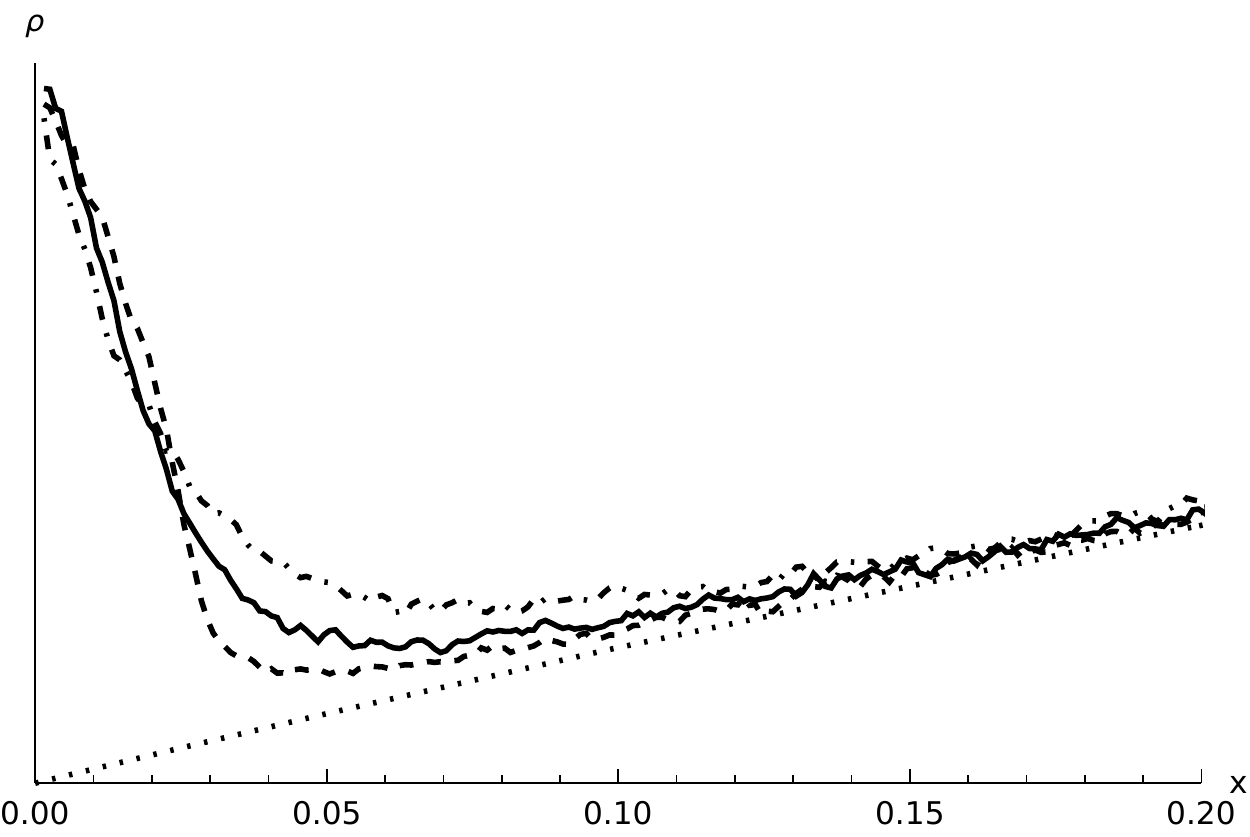}
	\caption{Comparison between three methods of resetting, overlaid on top of the Planck distribution (dotted straight line).  The three simulations have the same resetting rate  $r=0.01$, while the resetting parameters are $d=200$, $x_0\approx0.027$ and $\lambda_0\approx74$, all giving approximately equal first moment  $\langle x \rangle = 0.014...$. }\label{fig:compare}
\end{figure}

If, for the same given rate, the parameters are tuned such that these first moments coincide, we may expect the different resetting protocols to produce the same behavior on average whenever
\[\langle x \rangle_\text{Uni} = \langle x \rangle_\text{Exp}=\langle x\rangle_{\text{Div}} \implies \frac{x_0}{2}=\frac{1}{\lambda_0} = \frac{\langle x \rangle_\text{Planck}}{d}
\]
To verify that, we perform three simulations, one for each method, with a resetting rate of $r= \ $\num{0.01}, and the parameters chosen such that $\langle x \rangle$ after resetting equals $\approx 0.014...$. We indeed observe a reasonable agreement between the simulations with those different resetting-mechanisms, as seen in Fig.\ref{fig:compare}.\\

%\begin{figure}
%	\begin{subfigure}{0.48\textwidth}
%		\includegraphics[width=\textwidth]{spd_r001_d100.pdf}
%		\caption{$r=0.01$, $d=100$}
%	\end{subfigure}\hfill
%	\begin{subfigure}{0.48\textwidth}
%		\includegraphics[width=\textwidth]{spd_r001_d200.pdf}
%		\caption{$r=0.01$, $d=200$}
%	\end{subfigure}
%	
%	\begin{subfigure}{0.48\textwidth}
%		\includegraphics[width=\textwidth]{spd_r010_d100.pdf}
%		\caption{$r=0.10$, $d=200$}
%	\end{subfigure}\hfill
%	\begin{subfigure}{0.48\textwidth}
%		\includegraphics[width=\textwidth]{spd_r010_d200.pdf}
%		\caption{$r=0.10$, $d=200$}
%	\end{subfigure}
%	
%	\begin{subfigure}{0.48\textwidth}
%		\includegraphics[width=\textwidth]{spd_r050_d100.pdf}
%		\caption{$r=0.50$, $d=100$}
%	\end{subfigure}\hfill
%	\begin{subfigure}{0.48\textwidth}
%		\includegraphics[width=\textwidth]{spd_r050_d200.pdf}
%		\caption{$r=0.50$, $d=200$}
%	\end{subfigure}
%	\caption{Zooming in on low frequency for simulations with $10^7$ particles. The transition to the Planck law (dashed line) occurs for $x>0.2$ here.}\label{fig:gridplot}
%\end{figure}

\begin{figure}[th!]
	\includegraphics[width=0.85\textwidth]{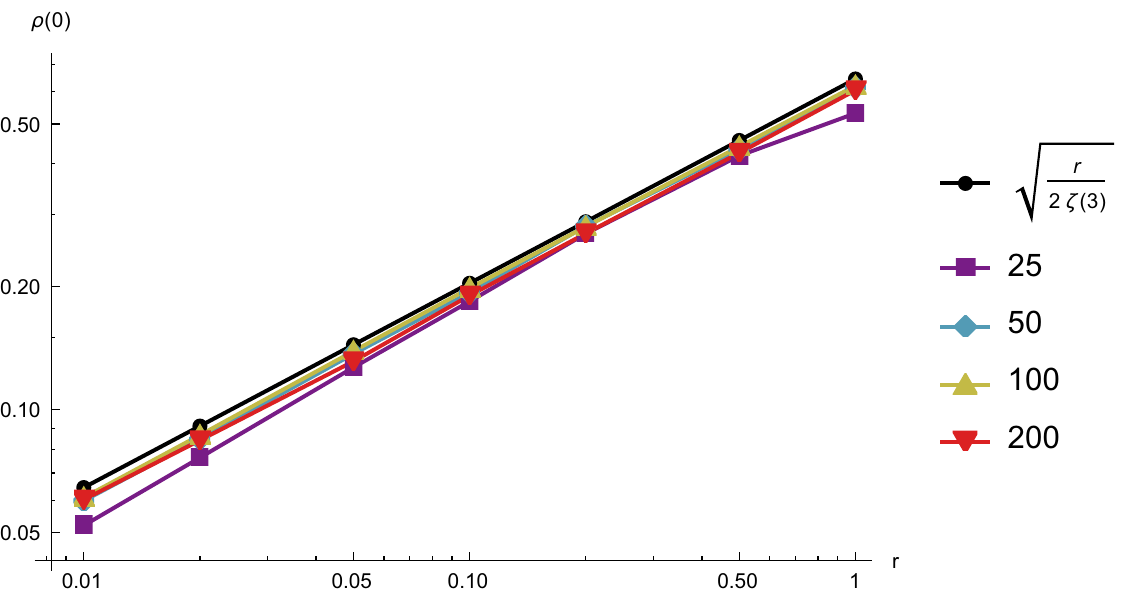}
	\caption{Spectral probability density at the origin in function of the rate $r$ for various divisors $d$.}\label{fig:near-divisors}
\end{figure}
%\begin{figure}
%	\includegraphics[width=\textwidth]{near-rates.pdf}
%	\caption{Condensate size in function of the divisor for various rates.  The decrease for divisor $d>500$ is due to the thermal push back to the Planck distribution when the frequencies gets too small. As before, $N_0$ is the number of photons with frequency $<0.1$. \textbf{WE DONOTNEED THE BLUE AND PURPLECURVES}}\label{fig:near-rates}
%\end{figure}

That motivates using only one resetting protocol, which is taken to be the ``division'' (Doppler shift). In order to explore the effects of the various parameters, we take $r=0.02,0.05,0.10,0.20$, varying the divisor $d$ from 25 to 400. Pictures of the frequency probability distribution at (very) low frequency are shown in Fig.\ref{fig:exploration}.\\ 
Notice there the presence of a non-zero amount of photons at frequency zero, rejoining the Planck distribution either gradually or with a strong jump downwards first.  To quantify the size of this condensate, we estimate the probability density at the origin by averaging over the 10 lowest bins (for comparison, this is 0.1\% of the support of the Planck distribution). We plot the result for various rates and divisors in Fig~\ref{fig:near-divisors} (log-log plots). The curves for different divisors overlap reasonably well with a power law having an exponent of 0.5. \\
Secondly, we observe that the intercept with the $y$-axis seems to be independent of the divisor $d$.  To understand that, we calculate the Kompaneets flux  $(2-x)x\rho(x) - x^2 \rho'(x) - 2\zeta(3) \rho(x)^2$, to be $-2\zeta(3)\rho(0)^2$ near the origin.  That has to match the resetting flux, which is simply $r$,  implying that $\rho(0) \approx \sqrt{\frac{r}{2\zeta(3)}}$.  That explains the exponent 0.5 in the curves of Fig.~\ref{fig:near-divisors}  and that prediction for the zero-frequency density is indeed also quantitatively close to what we observe, though it is not entirely accurate.

\section {Conclusions}\label{con}
We have applied a frequency-resetting mechanism to a tagged photon undergoing Compton scattering from a thermal matter environment.  The frequency is Doppler shifted at random moments by imagining a punctuated increase of the scale factor, e.g. in an FLRW-scenario with cosmic expansion ``in fits and starts'' applied to the primordial plasma before recombination. Alternatively, we may consider optomechanical setups where a cavity  randomly and  quickly expands. Strongly inelastic scattering with matter can also provide a source for resetting, but further studies on the exact mechanisms must be carried.\\
The result of this resetting is naturally felt mostly at low frequencies while the Planck law is almost perfectly preserved at moderate to high frequencies.  It may be an observable effect for certain parameters and we feel particularly encouraged by the preliminary results of observation and analysis of the ARCADE-data; see \cite{arca,arcade1,arcade2,edges}.  The jury is still out however whether deviations from the Planck law would be really there in the cosmic microwave background.  Whether quantum optical experiments (in our labs on Earth) as well can reproduce the obtained predictions is an open question and exciting challenge.

\bibliography{langevin-kompaneets}

\end{document}